\documentclass[aps,preprint]{revtex4}%
\usepackage{amsfonts}
\usepackage{amsmath}
\usepackage{amssymb}
\usepackage{graphicx}%
\begin{document}
\title[Short title for running header]{Nonextensive and superstatistical generalizations of random-matrix theory}
\author{A.Y. Abul-Magd}
\affiliation{Faculty of Engineering Sciences, Sinai University, El Arish, Egypt}
\keywords{Random-matrix theory, non-extensive entropy, superstatistics}
\pacs{05.40.-a, 05.45.Mt, 03.65.-w, 02.30.Mv}

\begin{abstract}
Random matrix theory (RMT) is based on two assumptions: (1) matrix-element
independence, and (2) base invariance. Most of the proposed generalizations
keep the first assumption and violate the second. Recently, several authors
presented other versions of the theory that keep base invariance on the
expense of allowing correlations between matrix elements. This is achieved by
starting from non-extensive entropies rather than the standard Shannon
entropy, or following the basic prescription of the recently suggested concept
of superstatistics. We review these generalizations of RMT and illustrate
their value by calculating the nearest-neighbor-spacing distributions and
comparing the results of calculation with experiments and
numerical-experiments on systems in transition from order to chaos

\end{abstract}
\date{\today}
\startpage{01}
\endpage{02}
\maketitle

\section{Introduction}

In classical mechanics, integrable Hamiltonian dynamics is characterized by
the existence of as many conserved quantities as degrees of freedom. Each
trajectory in the corresponding phase space evolves on an invariant
hyper-torus \cite{lichtenberg,chandre}. In contrast, chaotic systems are
ergodic; almost all orbits fill the energy shell in a uniform way. Physical
systems with integrable and fully chaotic dynamics are exceptional. A typical
Hamiltonian system shows a mixed phase space in which regions of regular
motion and chaotic dynamics coexist . Systems of this kind are known as mixed
systems. Their dynamical behavior is by no means universal. If we perturb an
integrable system, most of the periodic orbits on tori with rational
frequencies disappear. However, some of these orbits persist. Elliptic
periodic orbits appear surrounded by islands. They correspond to librational
motions around these periodic orbits and reflect their stability. The
Kolmogorov-Arnold (KAM) theorem establishes the stability with respect to
small perturbations of invariant tori with a sufficiently incommensurate
frequency vector. When the perturbation increases, numerical simulations show
that more and more tori are destroyed. For large enough perturbations, there
are locally no tori in the considered region of phase-space. The break-up of
invariant tori leads to a loss of stability of the system, to chaos. Different
scenaria of transition to chaos in dynamical systems have been considered.
There are three main scenaria of transition to global chaos in
finite-dimensional (non-extended) dynamical systems: via the cascade of
period-doubling bifurcations, the Lorenz system-like transition via Hopf and
Shil'nikov bifurcations, and the transition to chaos via intermittences
\cite{eckmann,elnashaie,bunimovich}. It is natural to expect that there could
be other (presumably many more) such scenaria in extended
(infinite-dimensional) dynamical systems.

In quantum mechanics, the specification of a wave function is always related
to a certain basis. In integrable systems eigenbasis of the Hamiltonian is
known in principle. In this basis, each eigenfunction has just one component
that obviously indicates the absence of complexity. In the nearly ordered
regime, mixing of quantum states belonging to adjacent levels can be ignored
and the energy levels are uncorrelated. The level-spacing distribution
function obeys the Poissonian, $\exp(-s)$, where $s$ is the energy spacing
between adjacent levels normalized by the mean level spacing $D$. On the other
hand, the eigenfunctions a Hamiltonian with a chaotic classical limit is
unknown in principle. In other words, there is no special basis to express the
eigenstates of a chaotic system. If we try to express the wave functions of a
chaotic system in terms of a given basis, their components become on average
uniformly distributed over the whole basis. They are also extended in all
other bases. For example, Berry \cite{berry} conjectured that the
wavefunctions of chaotic quantum systems can be represented as a formal sum
over elementary solutions of the Laplace equation in which real and imaginary
parts of coefficients are independent identically-distributed Gaussian random
variables with zero mean and variance computed from the normalization. Bohigas
et al. \cite{bohigas} put forward a conjecture (strongly supported by
accumulated numerical evidence) that the spectral statistics of chaotic
systems follow random-matrix theory (RMT) \cite{mehta,guhr}. This theory
models a chaotic system by an ensemble of random Hamiltonian matrices
$\mathbf{H}$ that belong to one of the three universal classes, orthogonal,
unitary and symplectic. The theory is based on two main assumptions: the
matrix elements are independent identically-distributed random variables, and
their distribution is invariant under unitary transformations. These lead to a
Gaussian probability density distribution for the matrix elements. The
Gaussian distribution is also obtained by maximizing the Shannon entropy under
constraints of normalization and existence of the expectation value of
Tr$\left(  \mathbf{H}^{\dagger}\mathbf{H}\right)  $, where Tr denotes the
trace and $\mathbf{H}^{\dagger}$ stands for the Hermitian conjugate of
$\mathbf{H}$ \cite{mehta,balian}. The statistical information about the
eigenvalues and/or eigenvectors of the matrix can be obtained by integrating
out all the undesired variables from distribution of the matrix elements. This
theory predicts a universal form of the spectral correlation functions
determined solely by some global symmetries of the system (time-reversal
invariance and value of the spin). Time-reversal-invariant quantum system are
represented by a Gaussian orthogonal ensemble (GOE) of random matrices when
the system has rotational symmetry and by a Gaussian symplectic ensemble (GSE)
otherwise. Chaotic systems without time reversal invariance are represented by
the Gaussian unitary ensemble (GUE). The dimension $\beta$ of the underlying
parameter space is used to label these three ensembles: for GOE, GUE and GSE,
$\beta$ takes the values 1, 2 and 4, respectively. Among several measures
representing spectral correlations, the nearest-neighbor level-spacing
distribution function $p(s)$ has been extensively studied so far. According to
the random matrix theory, the level spacing distribution function in the
chaotic phase is approximated by the Wigner-Dyson distribution, namely,%
\begin{equation}
P_{\beta}(s)=a_{\beta}s^{\beta}\exp(-b_{\beta}s^{2}).
\end{equation}
The coefficients $a_{\beta}$\ and $b_{\beta}$ are determined by the
normalization conditions $\int_{0}^{\infty}P_{\beta}(s)ds=\int_{0}^{\infty
}sP_{\beta}(s)ds=1$, as $a_{1}=\pi/2,~a_{2}=32/\pi^{2},~a_{4}=2^{18}/36\pi
^{3},~b_{1}=\pi/4,~b_{2}=4/\pi,$and$~b_{4}=64/9\pi$. For $s\ll1$, the
distribution function is proportional to $s^{\beta}$, which implies that
adjacent energy levels cannot approach each other indefinitely because of
mixing between two extended states.

The assumptions that lead to RMT do not apply for mixed systems. The
Hamiltonian of a typical mixed system can be described as a random matrix with
some (or all) of its elements as randomly distributed. Here the distributions
of various matrix elements need not be same, may or may not be correlated and
some of them can be non-random too. This is a difficult route to follow. So
far in the literature, there is no rigorous statistical description for the
transition from integrability to chaos. There have been several proposals for
phenomenological random matrix theories that interpolate between the
Wigner-Dyson RMT and banded RM with the (almost) Poissonian level statistics.
The standard route of the derivation is to sacrifice basis invariance but keep
matrix-element independence. The first work in this direction is due to
Rosenzweig and Porter \cite{rosen}. They model the Hamiltonian of the mixed
system by a superposition of two matrices: a diagonal matrix of random
elements having the same variance and a matrix drawn from a GOE. Therefore,
the variances of the diagonal elements total Hamiltonian are different from
those of the off-diagonal ones, unlike the standard GOE Hamiltonian in which
the variances of diagonal elements are twice of the off-diagonal ones. Hussein
and Sato \cite{hussein} used the maximum entropy principle to construct such
ensembles by imposing additional constraints. Ensembles of band random
matrices whose entries are equal to zero outside a band of width $b$ along the
principal diagonal have also been often used to model mixed systems
\cite{casati,fyodorov,mirlin,kravtsov,evers}.

Another route for generalizing RMT is to conserve base invariance but allow
for correlation of matrix elements. This has been achieved by maximizing
non-extensive entropies subject to the constraint of fixed expectation value
of Tr$\left(  H^{\dagger}H\right)  $
\cite{evans,toscano,nobre,abul,bertuola,abul1,abul2}. Recently, an equivalent
approach is presented in \cite{sust1,sust2}, which is based on the method of
superstatistics (statistics of a statistics) proposed by Beck and Cohen
\cite{BC}. This formalism has been elaborated and applied successfully to a
wide variety of physical problems, e.g., in
\cite{cohen,beck,beckL,salasnich,sattin,reynolds,ivanova,beckT}. In
thermostatics, superstatistics arises as weighted averages of ordinary
statistics (the Boltzmann factor) due to fluctuations of one or more intensive
parameter (e.g. the inverse temperature). Its application to RMT assumes the
spectrum of a mixed system is made up of many smaller cells that are
temporarily in a chaotic phase. Each cell is large enough to obey the
statistical requirements of RMT but has a different distribution parameter
$\eta$ associated with it, according to a probability density $f(\eta)$.
Consequently, the superstatistical random-matrix ensemble describes the mixed
system as a mixture of Gaussian ensembles with a statistical weight $f(\eta
).$Therefore one can evaluate any statistic for the superstatistical ensemble
by simply integrating the corresponding statistic for the conventional
Gaussian ensemble.

\section{Nonextensive generalization of RMT}

In 1957 Jaynes \cite{jaynes} proposed a rule, based on information theory, to
provide a constructive criterion for setting up probability distributions on
the basis of partial knowledge. This leads to a type of statistical inference
which is called the maximum-entropy principle (MaxEnt). It is the least biased
estimate possible on the given information. Jaynes showed in particular how
his rule, when applied to statistical mechanics, leads to the usual Gibbs'
canonical distribution. The core of the MaxEnt method resides in interpreting
entropy, through the Shannon axioms, as a measure of the \textquotedblleft
amount of uncertainty\textquotedblright\ or of the \textquotedblleft amount of
information that is missing\textquotedblright\ in a probability distribution.
This was an important step forward because it extended the applicability of
the notion of entropy far beyond its original roots in thermodynamics. Balian
considered the application of MaxEnt to the random-matrix theory by maximizing
the Shannon entropy under constraints of normalization and existence of the
expectation value of Tr$\left(  \mathbf{H}^{\dagger}\mathbf{H}\right)  $. In
this section, we consider possible generaliztion of RMT by extremizing two
different entropies, namely Tsallis' and Kaniadakis', rather than Shannon's
entropy. The extremization is again subject to the constraint of normalization
and existence of the expectation value of Tr$\left(  \mathbf{H}^{\dagger
}\mathbf{H}\right)  $. For completeness, we start by a brief review of the
conventional random-matrix theory.

\subsection{RMT from Shannon's entropy}

Balian \cite{balian} derived the weight functions $P(\mathbf{H})$ for the
random-matrix ensembles from MaxEnt postulating the existence of a second
moment of the Hamiltonian. He applied the conventional Shannon definition for
the entropy to ensembles of random matrices as%
\begin{equation}
S_{\text{Sch}}=-\int d\mathbf{H}P(\mathbf{H})\ln P(\mathbf{H})
\end{equation}
and maximized it under the constraints of normalization of $P(\mathbf{H}%
)$\ and fixed mean value of Tr$\left(  \mathbf{H}^{\dagger}\mathbf{H}\right)
$. The latter constraint ensures basis independence, which is a property of
the trace of a matrix. Then, the distribution $P(H)$ is determined from the
extremum of the functional%
\begin{equation}
F_{\text{Sch}}=S_{\text{Sch}}-\xi\int d\mathbf{H}~P(\mathbf{H})-\eta\int
d\mathbf{H}P(\mathbf{H})\text{Tr}\left(  H^{T}H\right)  ,
\end{equation}
where $\xi$ and $\eta$ are Lagrange multipliers. Its maximum is obtained
equating its functional derivative to zero. He obtained
\begin{equation}
P_{\beta}\left(  \mathbf{H}\right)  =\frac{1}{Z(\eta)}\exp\left[
-\eta\text{Tr}\left(  \mathbf{H}^{\dagger}\mathbf{H}\right)  \right]  ,
\end{equation}
where $Z(\eta)=\int\exp\left[  -\eta\text{Tr}\left(  \mathbf{H}^{\dagger
}\mathbf{H}\right)  \right]  d\mathbf{H}$.

It is easy to see that the joint distribution of matrix elements obtained in
Eq. (4) satisfies the two conditions of RMT, namely uncorrelated
matrix-elements and Base independence. The first condition follow since the
distribution (4) is a Gaussian distribution with inverse variance $1/2\eta$,
since Tr$\left(  \mathbf{H}^{\dagger}\mathbf{H}\right)  =\sum_{i=1}^{N}\left(
H_{ii}^{(0)}\right)  ^{2}+2\sum_{\gamma=0}^{\beta-1}\sum_{i>j}\left(
H_{ij}^{(\gamma)}\right)  ^{2}$, where all the four matrices $H^{(\gamma)}$
with $\gamma=0,1,2,3$ are real. This allows the factorization of $P_{\beta
}\left(  \mathbf{H}\right)  $ into products of terms depending only on the
individual matrix elements. Therefore, the matrix elements of $\mathbf{H}$ are
independent. Base independence follows from the fact that the distribution
(18) depends on $\mathbf{H}$ in the combination Tr$\left(  \mathbf{H}%
^{\dagger}\mathbf{H}\right)  $. Indeed, if two matrices \textbf{A} and
\textbf{B} that express the same operator in two different bases are related
by a similarity transformation \textbf{B} = $\mathbf{T}^{-1}\mathbf{AT}$, then
such operators have the same trace.

The joint distribution of eigenvalues $E_{i}$ immediately follows from Eq.
(3). With $\mathbf{H}=\mathbf{U}^{-1}\mathbf{EU}$, where $\mathbf{U}$\ is the
global unitary group and $\mathbf{E}$ $=$ diag$(E_{1},\cdots,E_{N})$ the
volume element $d\mathbf{H}$ has the form
\begin{equation}
d\mathbf{H}=\left\vert \Delta_{N}\left(  \mathbf{E}\right)  \right\vert
^{\beta}d\mathbf{E}d\mu(\mathbf{U}),
\end{equation}
where $\Delta_{N}\left(  \mathbf{E}\right)  =\prod_{n>m}(E_{n}-E_{m})$ is the
Vandermonde determinant and $d\mu(\mathbf{U})$ the invariant Haar measure of
the unitary group \cite{mehta,guhr}. Integrating over $\mathbf{U}$ \ and
noting that Tr$\left(  \mathbf{H}^{\dagger}\mathbf{H}\right)  =$%
Tr$\mathbf{E}^{2}$\ yields the joint probability density of eigenvalues in the
form%
\begin{equation}
P_{\beta}(\eta,E_{1},\cdots,E_{N})=C_{\beta}\prod_{n>m}\left\vert E_{n}%
-E_{m}\right\vert ^{\beta}\exp\left(  -\eta\sum_{i=1}^{N}E_{i}^{2}\right)  ,
\end{equation}
where $C_{\beta}$ is a normalization constant. All of the spectral properties
of the Gaussian random-matrix ensemble can be obtained from Eq. (6). However,
this is not an easy task. Lacking simple exact results, and guided by the case
$N$ = 2, Wigner proposed a form for the nearest neighbor spacing (NNS)
distribution $p(s)$ of eigenvalues. This \textquotedblleft Wigner
surmise\textquotedblright, originally stated for $\beta$ = 1, has the form%
\begin{equation}
p_{\beta}(s,\eta)=\frac{\sqrt{2\mu}}{\Gamma\left[  \left(  \beta+1\right)
/2\right]  }\left(  \frac{\eta s}{2}\right)  ^{\beta}\exp\left(  -\frac{\eta
s^{2}}{2}\right)  . \label{Wig}%
\end{equation}
The parameter $\eta$ is determined by the condition of unit mean spacing,
$\int_{0}^{\infty}s~p_{\beta}(s)ds=1$, as%
\begin{equation}
\eta=2\frac{\Gamma^{2}\left[  \left(  \beta+2\right)  /2\right]  }{\Gamma
^{2}\left[  \left(  \beta+1\right)  /2\right]  }.
\end{equation}
Although the Wigner surmise is strictly valid for two-dimensional ensembles,
it is an accurate approximation for ensemble with arbitrarily large $N$. To
demonstrate the accuracy, we expand this distribution for the case of
$\beta=2$ in powers of $s$ to obtain%
\begin{equation}
p_{2}(s)=\frac{32}{\pi^{2}}s^{2}\left(  1-\frac{4}{\pi}s^{2}+\cdots\right)
\cong3.242s^{2}-4.128s^{4}+\cdots,
\end{equation}
while the power-series expansion of the corresponding exact distribution for
ensembles with $N\rightarrow\infty$ \cite{mehta} yields%
\begin{equation}
p_{2,\text{exact}}(s)=\frac{\pi^{2}}{3}s^{2}-\frac{2\pi^{4}}{45}s^{4}%
+\cdots\cong3.290s^{2}-4.329s^{4}+\cdots.
\end{equation}
The Wigner surmise has been successfully applied to the NNS distributions for
numerous chaotic systems.

\subsection{RMT from Tsallis' entropy}

The past decade has witnessed a considerable interest devoted to
non-conventional statistical mechanics. Much work in this direction followed
the line initiated by Tsallis' seminal paper \cite{Ts1}. The standard
statistical mechanics is based on the Shannon entropy measure $S=-\Sigma
_{i}p_{i}\ln p_{i\text{ }}$(we use Boltzmann's constant $k_{B}=1$), where
$\{p_{i}\}$ denotes the probabilities of the microscopic configurations. This
entropy is extensive. For a composite system $A+B$, constituted of two
independent subsystems $A$ and $B$ such that the probability
$p(A+B)=p(A)p(B),$ the entropy of the total $S(A+B)=S(A)+S(B)$. Tsallis
proposed a non-extensive generalization: $S_{q}=\left(  1-\Sigma_{i}p_{i}%
^{q}\right)  /(q-1)$. The entropic index $q$ characterizes the degree of
extensivity of the system. The entropy of the composite system $A+B$, the
Tsallis' measure verifies
\begin{equation}
S_{q}(A+B)=S_{q}(A)+S_{q}(B)+(1-q)S_{q}(A)S_{q}(B),
\end{equation}
from which the denunciation non-extensive comes. Therefore, $S_{q}%
(A+B)<S_{q}(A)+S_{q}(B)$ if $q>1$. This case is called sub-extensive. If
$q<1$, the system is in the super-extensive regime. The standard statistical
mechanics recovered for $q$ = 1. Applications of the Tsallis formalism covered
a wide class of phenomena; for a review please see, e.g. \cite{Ts2}.

The Tsallis entropy is defined for the joint matrix-element probability
density $P_{\beta}(q,\mathbf{H})$\ by
\begin{equation}
S_{q}\left[  P_{\text{Ts,}\beta}(q,\mathbf{H})\right]  =\left.  \left(  1-\int
dH\left[  P_{\text{Ts,}\beta}(q,\mathbf{H})\right]  ^{q}\right)  \right/
(q-1).
\end{equation}
We shall refer to the corresponding ensembles as the Tsallis orthogonal
ensemble (TsOE), the Tsallis Unitary ensemble (TsUE), and the Tsallis
symplectic ensemble (TsSE). For $q\rightarrow1$, $S_{q}\ $tends to Shannon's
entropy, which yields the canonical Gaussian orthogonal, unitary or symplectic
ensembles\ (GOE, GUE, GSE) \cite{mehta,balian}.

There are more than one formulation of non-extensive statistics which mainly
differ in the definition of the averaging. Some of them are discussed in
\cite{wang}. We apply the most recent formulation \cite{Ts3}. The probability
distribution $P_{\text{Ts,}\beta}(q,\mathbf{H})$\ is obtained by maximizing
the entropy under two conditions, where $\sigma_{\beta}$ is a constant. The
optimization of $S_{q}$ with these constraints yields a power-law type for%
\begin{equation}
P_{\beta}(q,H)P_{\text{Ts,}\beta}(q,\mathbf{H})=\widetilde{Z}_{q}^{-1}\left[
1+(q-1)\widetilde{\eta}_{q}\left\{  \text{Tr}\left(  \mathbf{H}^{\dagger
}\mathbf{H}\right)  -\sigma_{\beta}^{2}\right\}  \right]  ^{-\frac{1}{q-1}},
\end{equation}
where $\widetilde{\eta}_{q}>0$ is related to the Lagrange multiplier $\eta$
associated with the constraint of fixed Tr$\left(  \mathbf{H}^{\dagger
}\mathbf{H}\right)  $ by
\begin{equation}
\widetilde{\eta}_{q}=\eta/\int d\mathbf{H}\left[  P_{\text{Ts,}\beta
}(q,\mathbf{H})\right]  ^{q},
\end{equation}
and\
\begin{equation}
\widetilde{Z}_{q}=\int d\mathbf{H}\left[  1+(q-1)\widetilde{\eta}_{q}\left\{
\text{Tr}\left(  \mathbf{H}^{\dagger}\mathbf{H}\right)  -\sigma_{\beta}%
^{2}\right\}  \right]  ^{-\frac{1}{q-1}}.
\end{equation}
It turns out that the distribution (13) can be written hiding the presence of
$\sigma_{\beta}^{2}$ in a more convenient form
\begin{equation}
P_{\text{Ts,}\beta}(q,\mathbf{H})=Z_{q}^{-1}\left[  1+(q-1)\eta_{q}%
\text{Tr}\left(  \mathbf{H}^{\dagger}\mathbf{H}\right)  \right]  ^{-\frac
{1}{q-1}},
\end{equation}
where
\begin{equation}
\eta_{q}=\frac{\eta}{\int d\mathbf{H}\left[  P_{\beta}(q,\mathbf{H})\right]
^{q}+(1-q)\eta\sigma_{\beta}^{2}},
\end{equation}
and
\begin{equation}
Z_{q}=\int d\mathbf{H}\left[  1+(q-1)\eta_{q}\text{Tr}\left(  \mathbf{H}%
^{\dagger}\mathbf{H}\right)  \right]  ^{-\frac{1}{q-1}}.
\end{equation}
The probability density $P_{\text{Ts,}\beta}(q,\mathbf{H})$ depends on
$\mathbf{H}$ through Tr$\left(  \mathbf{H}^{\dagger}\mathbf{H}\right)  $\ and
is therefore invariant under arbitrary rotations in the matrix space. This
ensures base invariance. It decays by a power law as the square of any matrix
element tends to infinity in contrast with the Gaussian decay of the
distribution function of the conventional random matrix ensembles.

We now calculate the joint probability density for the eigenvalues of the
Hamiltonian $\mathbf{H}$. Expressing the volume element in the matrix-element
space in the form (5) and integrating over the "angular" variables, on obtains%
\begin{equation}
P_{\text{Ts,}\beta}(\eta_{q},E_{1},\cdots,E_{N})=C_{\text{Ts,}\beta}%
\prod_{n>m}\left\vert E_{n}-E_{m}\right\vert ^{\beta}\left[  1+(q-1)\eta
_{q}\sum_{i=1}^{N}E_{i}^{2}\right]  ^{-\frac{1}{q-1}},
\end{equation}
where $C_{\text{Ts,}\beta}$ is a normalization constant.

In order to obtain a generalization of Wigner's surmise, we consider the
special case of $N=2.$ In this case,%
\begin{equation}
P_{\text{Ts,}\beta}(\eta_{q};\varepsilon,s)=c_{\text{Ts,}\beta}s^{\beta
}\left[  1+(q-1)\eta_{q}\left(  2\varepsilon^{2}+\frac{1}{2}s^{2}\right)
\right]  ^{-\frac{1}{q-1}}, \label{PTsE}%
\end{equation}
where $\varepsilon=(E_{1}+E_{2})/2,~s=\left\vert E_{1}-E_{2}\right\vert $. For
this case, the distribution (\ref{PTsE}) has to be complemented by the
auxiliary condition that the quantity inside the square bracket has to be
positive. We here consider the case of $q\geq1$ where no limitations are
imposed on the values of the variables $\varepsilon$ and $s$, and refer the
reader interested in the other case of $q<1$ to Ref. \cite{abul}. The NNS
distribution is obtained by integrating (\ref{PTsE}) over $\varepsilon$ from
$-\infty$ to $\infty$.%
\begin{equation}
p_{\text{Ts,}\beta}(q,s)=a_{\text{Ts,}\beta}s^{\beta}\left[  1+b_{\text{Ts,}%
\beta}s^{2}\right]  ^{-\frac{1}{q-1}+\frac{1}{2}}, \label{PSTs}%
\end{equation}
where $a_{\text{Ts,}\beta}$\ is a normalization coefficient and $b_{\text{Ts,}%
\beta}$\ is obtained for the requirement of unit mean spacing. Explicitly,%
\begin{equation}
a_{\text{Ts,}\beta}=\frac{2b_{\text{Ts,}\beta}^{\left(  \beta+1\right)
/2}~\Gamma\left(  \frac{1}{q-1}-\frac{1}{2}\right)  }{\Gamma\left(
\frac{\beta+1}{2}\right)  \Gamma\left(  \frac{1}{q-1}-\frac{\beta}%
{2}-1\right)  }\text{ \ and \ }b_{\text{Ts,}\beta}=\frac{\Gamma^{2}\left(
\frac{\beta+2}{2}\right)  \Gamma^{2}\left(  \frac{1}{q-1}-\frac{\beta}%
{2}-\frac{3}{2}\right)  }{\Gamma^{2}\left(  \frac{\beta+1}{2}\right)
\Gamma^{2}\left(  \frac{1}{q-1}-\frac{\beta}{2}-1\right)  }%
\end{equation}
The second moment of \ the distribution $\left\langle s^{2}\right\rangle
=\int_{0}^{\infty}s^{2}p_{\beta}(s)ds$\ is given by%
\begin{equation}
\left\langle s^{2}\right\rangle =\frac{\Gamma\left(  \frac{\beta+1}{2}\right)
\Gamma\left(  \frac{\beta+3}{2}\right)  \Gamma\left(  \frac{1}{q-1}%
-\frac{\beta}{2}-1\right)  \Gamma\left(  \frac{1}{q-1}-\frac{\beta}%
{2}-2\right)  }{\Gamma^{2}\left(  \frac{\beta+2}{2}\right)  \Gamma^{2}\left(
\frac{1}{q-1}-\frac{\beta}{2}-\frac{3}{2}\right)  }%
\end{equation}
It diverges unless $q<q_{\infty}=1+2/\left(  \beta+4\right)  $, which is equal
to 1.40, 1.33 and 1$.$25 for the orthogonal, unitary and symplectic ensemble,
respectively. This imposes physical bound on the admissible values of $q$,
because $\left\langle s^{2}\right\rangle $ has to be finite in order to force
condition that Tr$\left(  \mathbf{H}^{\dagger}\mathbf{H}\right)  $\ has to be
finite. At higher values of the entropic index, non-extensive statistics does
not apply to the random matrix model. The peak of the distribution in Eq.
(\ref{PSTs}) is located at $s_{\beta}=\sqrt{\beta/b_{\text{Ts,}\beta}\left[
-1-\beta+1/(1-q)\right]  }\,$. It moves from $s_{1}$ = 0.798 to $s_{1}$ =
0.368, from $s_{2}$ = 0.886 to $s_{2}$ = 0.408 and from $s_{4}$ = 0.940 to
$s_{4}$ = 0.671 as $q=1$ (the standard Wigner's surmise) to $q_{\infty}$.
Neither reaches 0, the peak position of the Poisson distribution exp($-s$) of
the integrable systems. The proposed non-extensive ensemble in the three cases
of $\beta=1,2$ and 4 evolve the shape predicted by the corresponding Wigner
surmise towards the Poisson distribution, but never reach it.

\subsection{RMT from Kaniadakis' entropy}

In this section, we consider a possible generalization of RMT based on an
extremization of Kaniadakis' $\kappa$-entropy
\cite{kaniadakis,kaniadakis1,kaniadakis2}. This entropy shares the same
symmetry group of the relativistic momentum transformation and has
applications in cosmic-ray and plasma physics. For the matrix-element
probability distribution function, it reads%
\begin{equation}
S_{\kappa}\left[  \kappa,P_{\text{K,}\beta}(\kappa,\mathbf{H})\right]
=-\frac{1}{2\kappa}\int dH\left(  \frac{\alpha^{\kappa}}{1+\kappa}\left[
P_{\text{K,}\beta}(\kappa,\mathbf{H})\right]  ^{1+\kappa}-\frac{\alpha
^{-\kappa}}{1-\kappa}\left[  P_{\text{K,}\beta}(\kappa,\mathbf{H})\right]
^{1-\kappa}\right)
\end{equation}
with $\kappa$ a parameter with value between 0 and 1; the case of $\kappa=0$
corresponds to the Schannon entropy. Here, $\alpha$ is a real positive
parameter. Kaniadakis has considered two choices of $\alpha$, namely
$\alpha=1$ and $\alpha=Z$, where $Z$ is the generalized partition function. We
here adopt the second choice. The matrix-element distribution $P_{\text{K,}%
\beta}(\kappa,\mathbf{H})$ is obtained by extremizing the functional%
\begin{equation}
F_{\text{K}}=S_{\kappa}-\eta_{\text{K}}\int dH~P_{\text{K,}\beta}%
(\kappa,\mathbf{H})\text{Tr}\left(  \mathbf{H}^{\dagger}\mathbf{H}\right)  ,
\end{equation}
where $\eta_{\text{K}}$ is a Lagrange multiplier. One arrives to the following
distribution%
\begin{equation}
P_{\text{K,}\beta}(\kappa,\mathbf{H})=\frac{1}{Z_{\kappa}}\exp_{\left\{
\kappa\right\}  }\left[  -\eta_{\text{K}}\text{Tr}\left(  \mathbf{H}^{\dagger
}\mathbf{H}\right)  \right]  ,
\end{equation}
where%
\begin{equation}
Z_{\kappa}=\int d\mathbf{H~}\exp_{\left\{  \kappa\right\}  }\left[
-\eta_{\text{K}}\text{Tr}\left(  \mathbf{H}^{\dagger}\mathbf{H}\right)
\right]  .
\end{equation}
Here $\exp_{\left\{  \kappa\right\}  }\left[  x\right]  $ is the $\kappa
$-deformed exponential \cite{kaniadakis} which is defined by
\begin{equation}
\exp_{\left\{  \kappa\right\}  }\left[  x\right]  =\left(  \sqrt{1+\kappa
^{2}x^{2}}+\kappa x\right)  ^{1/\kappa}=\exp\left(  \frac{1}{\kappa
}\text{arcsinh }\kappa x\right)  .
\end{equation}
Concerning the asymptotic behavior of $P_{\text{K,}\beta}(\kappa,\mathbf{H})$
we easily obtain that%
\begin{equation}
P_{\text{K,}\beta}(\kappa,\mathbf{H})\sim\left\vert \kappa\eta_{\text{K}%
}\text{Tr}\left(  \mathbf{H}^{\dagger}\mathbf{H}\right)  \right\vert
^{-1/\left\vert \kappa\right\vert }%
\end{equation}
as the square of any of the matrix elements tends to infinity

The probability density $P_{\text{K}}(\kappa,\mathbf{H})$ depends on
$\mathbf{H}$ through Tr$\left(  \mathbf{H}^{\dagger}\mathbf{H}\right)  $\ and
is therefore invariant under arbitrary rotations in the matrix space. \ Using
Eq. (26) and integrating over the "angular variables" $\mathbf{U}$ yields the
joint probability density of eigenvalues in the form%
\begin{equation}
P_{\text{K,}\beta}(\kappa;E_{1},...,E_{N})=C_{\text{K,}\beta}\prod
_{n>m}\left\vert E_{n}-E_{m}\right\vert ^{\beta}\exp_{\left\{  \kappa\right\}
}\left[  -\eta_{\text{K}}\sum_{i=1}^{N}E_{i}^{2}\right]  . \label{PkappaE}%
\end{equation}
where $C_{\text{K,}\beta}$ is a normalization constant.

In order to obtain a generalization of the Wigner surmise, we consider the
case of two-dimensional random-matrix ensemble where $N=2$ and $n=3$ and
restrict our consideration to the orthogonal ensemble with $\beta=1.$ In this
case, Eq. (\ref{PkappaE}) reads%
\begin{equation}
P_{\text{K,}\beta}(\kappa;\varepsilon,s)=\frac{2\left(  1+3\kappa/4\right)
}{B\left(  \frac{1}{2\kappa}-\frac{3}{4},\frac{3}{2}\right)  }\left(
\kappa\eta_{\text{K}}\right)  ^{3/2}s\exp_{\left\{  \kappa\right\}  }\left[
-\eta_{\text{K}}\left(  2\varepsilon^{2}+\frac{1}{2}s^{2}\right)  \right]  ,
\label{PKE}%
\end{equation}
where $\varepsilon=(E_{1}+E_{2})/2,~s=\left\vert E_{1}-E_{2}\right\vert $, and
$B(a,b)=\Gamma(a)\Gamma(b)/\Gamma(a+b)$ is the Beta function \cite{gradshteyn}%
. The NNS distribution is obtained by integrating (\ref{PKE}) over
$\varepsilon$ from $-\infty$ to $\infty$. This can be done by changing the
variable $\varepsilon$ into $x=\exp[-\frac{1}{\kappa}$arcsinh$(\kappa
\eta_{\text{K}}s^{2}/2)]$, integrating by parts, and then replacing the
variable $x$ by another new variable, $y=\exp(\kappa x)$. The resulting
integral can be solved by using the following identity \cite{gradshteyn}%
\begin{multline}
\int_{u}^{\infty}y^{-\lambda}(y+\beta)^{\nu}\left(  y-u\right)  ^{\mu
-1}dy=u^{\mu+\nu-\lambda}B\left(  \lambda-\mu-\nu,\mu\right) \\
~{}_{2}F_{1}\left(  -\nu,\lambda-\mu-\nu\,;\lambda-\nu;-\frac{\beta}%
{u}\right)  ,
\end{multline}
for $\left\vert \beta/u\right\vert \,<1$ and $0<\mu<\lambda-\nu$,\ where
$_{2}F_{1}(\nu,\mu\,;\lambda;x)$ is the hypergeometric function. Thus, after
straightforward calculations we can express the NNS\ distribution as
\begin{multline}
p_{\text{K,1}}(\kappa,s)=-2\left(  1+\frac{3}{4}\kappa\right)  \eta_{\text{K}%
}se^{\left(  1/2-1/\kappa\right)  \text{arcsinh}(\kappa\eta_{\text{K}}%
s^{2}/2)}\frac{B\left(  \frac{1}{\kappa}-\frac{1}{2},\frac{3}{2}\right)
}{B\left(  \frac{1}{2\kappa}-\frac{3}{4},\frac{3}{2}\right)  }{}\label{PSK}\\
_{2}F_{1}\left(  -\frac{1}{2},\frac{1}{\kappa}-\frac{1}{2}\,;\frac{1}{\kappa
}+1;-e^{-2\text{arcsinh}(\kappa\eta_{\text{K}}s^{2}/2)}\right)  .
\end{multline}
The condition of unit mean spacing defines the quantity $\eta_{\text{K}}$\ as%
\begin{equation}
\eta_{\text{K}}=\left[  \frac{\pi k^{3/2}\left(  1+\frac{3}{4}\kappa\right)
}{\left(  1-\kappa^{2}\right)  B\left(  \frac{1}{2\kappa}-\frac{3}{4},\frac
{3}{2}\right)  }\right]  ^{2}.
\end{equation}
We note that the function $B\left(  \frac{1}{\kappa}-\frac{1}{2},\frac{3}%
{2}\right)  $ diverges at $\kappa=\kappa_{c}=1/2$, which serves as an upper
limit for the range of variation of $\kappa$. We also note that the mean
square spacing diverges at $\kappa=\kappa_{\infty}=2/5$.

\section{Superstatistical generalization of RMT}

Let us first recall the basic idea underlying superstatistics. We will then
proceed to construct a generalization of RMT in the spirit of superstatistics.

\subsection{Beck and Cohen's superstatistics}

Consider a complex system in a nonequilibrium stationary state. Such a system
will be, in general, inhomogeneous in both space and time. Effectively, it may
be thought to consist of many spatial cells, in each of which there may be a
different value of some relevant intensive parameter, e.g. the inverse
temperature $\beta$. Beck and cohen \cite{BC} assumed that this quantity
fluctuates adiabatically slowly, namely that the time scale is much larger
than the relaxation time for reaching local equilibrium. In that case, the
distribution function of the non-equilibrium system consists in Boltzmann
factors $\exp\left(  -\beta H\right)  $ that are averaged over the various
fluctuating inverse temperatures%
\begin{equation}
F\left(  H\right)  =\int g(\beta)\frac{\exp(-\beta H)}{z(\beta)}%
d\beta\label{sust}%
\end{equation}
where $z(\beta)$ is a normalizing constant, and $g(\beta)$ is the probability
distribution of $\beta$. Let us stress that $\beta^{-1}$ is a local variance
parameter of a suitable observable, the Hamiltonian of the complex system in
this case. Ordinary statistical mechanics are recovered in the limit
$g(\beta)\rightarrow\delta\left(  \beta\right)  $. In contrast, different
choices for the statistics of may lead to a large variety of probability
distributions $F(H)$. Several forms for $g(\beta)$ have been studied in the
literature, e.g. \cite{BC,sattin,bcs}. In particularly, Tsallis statistics is
generated from Eq. (\ref{sust}) if $\beta$ is a chi-squared random variable%
\begin{equation}
g(\beta)=\frac{1}{\Gamma(\nu/2)}\left(  \frac{\nu}{2\beta_{0}}\right)
^{\nu/2}\beta^{\nu/2-1}e^{-\nu\beta/2\beta_{0}} \label{P1}%
\end{equation}
where $\beta_{0}$ is the mean value. A chi-squared distribution derives from
the summation of squares of $\nu$ Gaussian random variables $X_{1}$, . . . ,
$X_{\nu}$ due to various relevant degrees of freedom in the system, where the
$X_{i~}$are independent, and \ $\left\langle X_{i~}\right\rangle ~$= 0. If
$\beta^{-1}$ rather than $\beta$ is the sum of several squared Gaussian random
variables, the resulting distribution $g(\eta)$ is the inverse $\chi^{2}$
distribution given by%
\begin{equation}
g(\beta)=\frac{\beta_{0}}{\Gamma(\nu/2)}\left(  \frac{\nu\beta_{0}}{2}\right)
^{\nu/2~-\nu/2-2}\beta e^{-\nu\beta_{0}/2\beta}, \label{P2}%
\end{equation}
where again $\beta_{0}$ is the average of $\beta$. Instead of being a sum of
many contributions, the random variable $\beta$ may be generated by
multiplicative random processes. Then $\ln\beta=\sum_{i=1}^{\nu}\ln X_{i}$ is
a sum of Gaussian random variables. Thus it is log-normally distributed,
\begin{equation}
g(\beta)=\frac{1}{\sqrt{2\pi}v\beta}e^{-\left.  \left[  \ln(\beta/\mu)\right]
^{2}\right/  2v^{2}}, \label{P3}%
\end{equation}
which has an average $\mu\sqrt{w}$ and variance $\mu^{2}w(w-1)$, where
$w=\exp(v^{2})$.

\subsection{RMT within superstatistics}

To apply the concept of superstatistics to RMT, one assumes the spectrum of a
(mixed) system as made up of many smaller cells that are temporarily in a
chaotic phase. Each cell is large enough to obey the statistical requirements
of RMT but is associated with a different distribution of the parameter $\eta$
in Eq. (4) according to a probability density $f(\eta)$. Consequently, the
superstatistical random-matrix ensemble used for the description of a mixed
system consists of a superposition of Gaussian ensembles. Its joint
probability density distribution of the matrix elements is obtained by
integrating the distribution given in Eq.~(4) over all positive values of
$\eta$\ with a statistical weight $f(\eta)$,
\begin{equation}
P(H)=\int_{0}^{\infty}f(\eta)\frac{\exp\left[  -\eta\text{Tr}\left(
H^{\dagger}H\right)  \right]  }{Z(\eta)}d\eta. \label{PH}%
\end{equation}
Despite the fact that it is hard to make this picture rigorous, there is
indeed a representation which comes close to this idea \cite{caer,muttalib}.

Beck, Cohen and Swinney \cite{bcs} proposed the derivation of superstatistics
starting from time-series. The idea is that superstatistical thermostatics
results as a convolution of two statistics, one characterized by the Boltzmann
factor and the other corresponding to inverse-temperature fluctuations. This
requires the existence of two relaxation times. A justification for the use of
the above-mentioned superstatistical generalization of RMT in the study of
mixed systems, is given in \cite{DA}. It is based on the representation of
their energy spectra in the form of discrete time series in which the level
order plays the role of time. The representation of the suitably transformed
eigenvalues of a quantum system as a time series has recently allowed to
determine the degree of chaoticity of the dynamics of the system
\cite{relano,gomez,santhanam,manimaran,santhanam1}. Reference \cite{DA}
considers two billiards with mushroom-shaped boundaries as representatives of
systems with mixed regular--chaotic dynamics and three with the shape of
Lima\c{c}on billiards, one of them of chaotic and two of mixed dynamics. The
quantum eigenvalues and statistical properties of the eigenfunctions were
obtained experimentally by exploiting the equivalence of the Schr{\"{o}}dinger
equation of a plane quantum billiard and the Helmholtz equation for the
electric field strength in a cylindrical microwave resonator for wave lengths
longer than twice the height of the resonator. The billiards with mixed
dynamics have classical phase spaces of different structures for the two
families of billiards. The "time-series" analysis of their spectra indeed
manifests the existence of two relaxation lengths. The short one, which is
defined as the average length over which energy fluctuations are correlated,
is of the order of the mean level spacing. The long one, which is by an order
of magnitude larger, characterizes the typical linear size of the
heterogeneous domains of the total spectrum.

The new framework of RMT provided by superstatistics should now be clear. The
local mean spacing is no longer uniformly set to unity but allowed to take
different (random) values at different parts of the spectrum. The parameter
$\eta$ is no longer a fixed parameter but it is a stochastic variable with
probability distribution $f(\eta)$. The observed mean level spacing is just
the expectation value of the local ones. The fluctuation of the local mean
spacing is due to the correlation of the matrix elements, which disappears for
chaotic systems. In the absence of these fluctuations, $f(\eta)=\delta
(\eta-\eta_{0})$ and we obtain the standard RMT. Within the superstatistics
framework, we can express\ any statistic $\sigma(E)$ of a mixed system that
can in principle be\ obtained from the joint eigenvalue distribution by
integration over some of the eigenvalues, in terms of the corresponding
statistic $\sigma^{(G)}(E,\eta)$ for a Gaussian random ensemble. The
superstatistical generalization is given by
\begin{equation}
\sigma(E)=\int_{0}^{\infty}f(\eta)\sigma^{(G)}(E,\eta)d\eta. \label{PSig}%
\end{equation}
The remaining task of superstatistics is the computation of the distribution
$f(\eta)$, which has been introduced in Eq.~(\ref{PH}). The time series
analysis in Ref. \cite{DA} allows to derive the sape the parameter
distribution $f(\eta)$. The obtained distribution agrees better with the
inverse $\chi^{2}$ distribution given by Eq. (\ref{P2}) rather than the other
two distributions (\ref{P1}) and (\ref{P3}). We have already mentioned that
the $\chi^{2}$ distribution of the superstatistical parameter $\eta$ yields
Tsallis statistics for RMT, which is considered in the previous section. The
log-normal distribution does not lead to simple analytical results for the
important level statistics like the NNS distribution. For these reasons, we
shall confine our further consideration to the case of inverse $\chi^{2}$
distributed superstatistical parameter $\eta$.

\subsection{Superstatistical generalization of Wigner's surmise}

The superstatistics generalization of NNS distribution for an ensemble
belonging to a given symmetry class is obtained by substituting the NNS
distribution of the corresponding Gaussian ensemble $p_{\beta}(s,\eta)$ for
$\sigma^{(G)}(E,\eta)$\ in (\ref{PSig}) and integrating over the local mean
level spacing $\eta$
\begin{equation}
p_{\text{SS,}\beta}(s)=\int_{0}^{\infty}f(v)p_{\beta}(s,\eta)d\eta.
\label{Psust}%
\end{equation}
For an inverse $\chi^{2}$ distribution of $\eta$, given by Eq.~(\ref{P2}), one
obtains the following superstatistical NNS distribution
\begin{equation}
p_{\text{SS,}\beta}(\nu,s)=\frac{4\sqrt{\eta_{0}/\nu}}{\Gamma\left(  \frac
{\nu}{2}\right)  \Gamma\left(  \frac{1+\beta}{2}\right)  }\left(  \sqrt
{\eta_{0}\nu}s/2\right)  ^{\frac{1+\nu+\beta}{2}}K_{\frac{1+\nu-\beta}{2}%
}\left(  \sqrt{\eta_{0}\nu}s\right)  , \label{PS2}%
\end{equation}
where $K_{m}(x)$ is a modified Bessel function \cite{gradshteyn} and $\eta
_{0}$ again is determined by the requirement that the mean-level spacing
$\left\langle s\right\rangle $ equals unity,
\begin{equation}
\eta_{0}=\frac{16\pi}{\nu^{3}}\left[  \frac{\Gamma\left(  \frac{3+\nu}%
{2}\right)  \Gamma\left(  1+\frac{\beta}{2}\right)  }{\Gamma\left(  \frac{\nu
}{2}\right)  \Gamma\left(  \frac{1+\beta}{2}\right)  }\right]  ^{2}.
\end{equation}

The inverse $\chi^{2}$ distribution of $\eta$ follows when the quantity
$\eta^{-1}$\ is the sum of $\nu$ squared Gaussian random variables. If we take
this assumption literarily, we must restrict $\nu$ to take positive integer
values. As the transition from integrability to chaos is known to proceed
continuously, we have to relax this condition and allow $\nu$ to take any real
value between 1 and $\infty$. Let us restrict our following consideration to
the case of orthogonal symmetry with $\beta=1.$ Using the asymptotic
expression of the modified Bessel function \cite{gradshteyn}, we easily find%
\begin{equation}
\lim_{\nu\rightarrow\infty}p_{\text{SS,}1}(\nu,s)=\frac{\pi}{2}se^{-\pi
s^{2}/4},
\end{equation}
which is the Wigner surmise, as required. The other limit of $\nu\rightarrow1$
yields the semi-Poisson distribution%
\begin{equation}
p_{\text{SemiPoisson}}(s)=4se^{-2s},
\end{equation}
which is known to provide a satisfactory description for the spectra of
pseudointegrable systems such as planar polygonal billiards, when all their
angles are rational with $\pi$ \cite{bog}. The motion of the corresponding
classical systems in phase space is not restricted to a torus like for
integrable systems, but to a surface with a more complicated topology
\cite{richens}. We therefore conclude that the assumption that the inverse
square of the variance of matrix elements as an inverse $\chi^{2}$ variable
allows one to model the transition out of chaos (corresponding to $\nu\gg1)$
until the system reaches the state of quasi-integrability as the effective
number $\nu$ of components of $\eta^{-1}$ approaches 1. If one allows $\nu$ to
take lower values, then the distribution $\left(  \ref{PS2}\right)  $ tends to
the Poisson distribution as $\nu\rightarrow-1$;%
\begin{equation}
p_{\text{SS,}1}(-1,s)=e^{-s}.
\end{equation}
We there conclude that formula $\left(  \ref{PS2}\right)  $ can provide a
model for describing the stochastic transition all the way from integrability
to chaos passing by the stage of quasi-integrability.

\section{Comparison with numerical experiment}

The NNS distributions $p_{\text{Ts,}\beta}(q,s)$ and $p_{\text{K,}\beta
}(\kappa,s)$\ obtained above when the entropy is given by the Tsallis and
Kaniadakis entropies, respectively, as well as the superstatistical
distribution $p_{\text{SS,}\beta}(\nu,s)$ describe evolution of the spacing
distribution from the Wigner shape to the Poissonian. They can be useful for
describing systems with mixed regular-chaotic at least when more familiar
distributions such as Berry and Robnik's or Brody's distribution
\cite{berryrobnik,brody} do not fit the data satisfactorily. We shall
demonstrate this by using these distributions to fit the NNS distribution of
levels of a number of mixed systems.

\subsection{Mushroom billiards}

Billiards can be used as simple models in the study of Hamiltonian systems.
They consist of a point particle which is confined to a container of some
shape and reflected elastically on impact with the boundary. The shape
determines whether the dynamics inside the billiard is regular, chaotic or
mixed. The best-known examples of chaotic billiards are the Sinai billiard (a
square table with a circular barrier at its center) and the Bunimovich stadium
(a rectangle with two circular caps) \cite{bunimovichS}. Neighboring parallel
orbits diverge when they collide with dispersing components of the billiard
boundary. In mixed billiards, some neighboring parallel orbits converge at
first, but divergence prevails over convergence on average. Divergence and
convergence are balanced in integrable billiards such as circles and ellipses.\ 

Recently Bunimovich introduced the so-called `mushroom' billiard
\cite{bunimovichM} with the novel feature of a well-understood divided
phase-space comprising a single integrable region and a single ergodic one. We
restrict ourselves here to mushroom billiards which consist of a semicircular
region, the `hat' and a `stem', which is symmetrically attached to its base.
As the width of the stem varies from zero to the diameter of the hat, there is
a continuous transition from integrability (the semicircle billiard) to
ergodicity (in case of a rectangular stem the stadium billiard). In mushroom
billiards, the regular region has a well-defined semicircular border. It is
composed of those trajectories in the hat that never cross this border and
therefore remain in the hat forever. Their integrability is due to the
conservation of the reflection angle for collisions with the semicircular
boundary. The chaotic component consists of trajectories that enter the stem
of the mushroom billiard. Two mushroom billiards have been recently
investigated experimentally \cite{friedrich}. The ratio of the width of the
stem to the diameter of the hat is 1:3 for the small mushroom billiard and 2:3
for the large. Both billiards have mixed dynamics with classical phase spaces
of different structures for the two billiards. The degree of chaos, which is
the measure of all chaotic parts of the phase space, is 45.5~$\%$ and
82.9~$\%$, respectively. Both systems have been studied experimentally in the
quantum limit exploiting the analogy between a quantum billiard and a flat
cylindric microwave billiard of the same shape. The electromagnetic resonances
in each microwave cavity can directly be associated with quantum states in the
corresponding quantum billiard. For the evaluation of statistical measures, a
sufficiently large number of resonances is needed. The first 780 resonances
could be detected in the small billiards and 938 in the large one. The quantum
eigenvalues were obtained experimentally by exploiting the equivalence of the
Schr{\"{o}}dinger equation of a plane quantum billiard and the Helmholtz
equation for the electric field strength in a cylindrical microwave resonator
for wave lengths longer than twice the height of the resonator. To compare the
statistical properties of the eigenvalues with universal predictions
considered in the present paper, they have to be rescaled to unit mean
spacing. This is done by an unfolding procedure using Weyl's formula
\cite{Weyl}, which relates the billiard area and circumference to the number
of resonance frequencies below a given one.

We compared the resulting NNS distributions given in Eqs.~(\ref{PSTs}),
(\ref{PSK}) and (\ref{PS2}) with the experimental ones for the mushroom
billiards and the two Lima\c{c}on billiards with mixed dynamics. In Fig.~1 the
experimental results for $p_{1}(s)$ are shown by histogram together while the
distributions obtained by starting with the Tsallis and Kaniadakis entropies
and the superstatistical distributions are shown by the dashed, dashed-dotted
and solid lines, respectively. The best fit values of the parameters are given
in Table I, together with the $\chi^{2}$ values calculated by%
\begin{equation}
\chi^{2}=\frac{1}{N}\sum_{i=1}^{N}\left[  p_{1}\left(  s_{i}\right)
-p_{\text{X},1}\left(  s_{i}\right)  \right]  ^{2},
\end{equation}
where $N$ is the number of experimental spacings and X stands for Ts
(Tsallis), K (Kaniadakis) or SS (superstatistics).

\subsection{Random binary networks}

As another example of mixed systems, we consider a numerical experiment by Gu
et al. \cite{gu} on a random binary network. Impurity bonds are employed to
replace the bonds in an otherwise homogeneous network. In such a network,
there exist a lot of geometric resonances randomly distributing in the
resonant area. Based on the Green's-function formalism, the eigenvalues of
Green's-matrix are solved, the sequence of which forms the resonance spectrum
The authors of Ref. \cite{gu} numerically calculated more than 700 resonances
for each sample. For each impurity concentration $p$, they considered 1000
samples with totally more than 700 000 levels computed. Their results for four
values of concentration $p$ are compared with both the Tsallis, Kaniadakis and
superstatistical NNS distributions in Fig. 2. The best fit values of the
parameters as well as the corresponding $\chi^{2}$ value are given in Table
II. The high statistical significance of the data allows us to assume the
advantage of the superstatistical distributions for describing the results of
this experiment, as compared to the other two distribution families.

\section{Conclusion}

Random matrix theory is the statistical theory of random matrices, whose
entries fluctuate as independent Gaussian random numbers. The matrix-element
distribution is obtained by extremizing Shannon's entropy subject to the
constraint of normalization and constant trace of the square of the matrix.
The latter constraint renders the matrix-element distribution base
independent. While most of the previously proposed generaliztion of RMT
violate base invariance, the ones reviewed in this paper conserve it.
Non-extensive generalizations extremize nonextensive entropies such as
Tsallis' or Kaniadakis', rather than Shannon's. Superstatistical
generalizations, on the other hand, allow the fluctuation of the mean local
density of states which is fixed in the standard theory. These generalizations
of RMT, seen from the present perspective, may bear interest per se because of
the additional nontrivial fluctuations introduced in a simple model. In
addition, they may constitute a useful statistical paradigm for the analysis
of the spectral fluctuations of systems with mixed regular-chaotic dynamics.
For this purpose, simple analytical expressions are derived in each case for
the nearest neighbor level distributions, being among the most popular
characteristics of spectral fluctuation. The formalism has been checked by the
analysis of experimental resonance spectra of mixed microwave billiards and
geometrical resonances in random binary networks. The predictions of the three
models satisfactorily describe the experimental trends of the evolution of NNS
distributions during the transition out of chaos. The considered experimental
data agree better in most cases with the corresponding distributions predicted
by the superstatistical approach when the fluctuating parameter has an inverse
$\chi^{2}$ distribution.

\pagebreak

TABLE I. Best-fit parameters for the experimental NNS distribution of
resonances in the small and large mushroom billiards. The corresponding
$\chi^{2}$ values are given in parentheses.%

\begin{tabular}
[c]{lll}%
Distribution & Small billiard & Large billiard\\
Tsallis & $q~$= 1.336 (0.0189) & $q~$= 1.221 (0.0031)\\
Kaniadakis & $\kappa$ = 0.423 (0.0159) & $\kappa$ = 0.017 (0.0069)\\
Superstatistical & $\nu$ = -0.441 (0.0026) & $\nu~$= 2.31 (0.0018)
\end{tabular}

\bigskip

TABLE II. Best-fit parameters for the\ numericcal-experimental NNS
distribution of geometrical resonances in the binary random network with
different impurity concentrations $p$. The corresponding $\chi^{2}$ values are
given in parentheses.%

\begin{tabular}
[c]{lllll}%
Distribution & $p=0.1$ & $p=0.2$ & $p=0.3$ & $p=0.4$\\
Tsallis & $q~$= 1.380 (0.0034) & $q~$= 1.322 (0.0022) & $q~$= 1.263 (0.0015) &
$q~$= 1.219 (0.0009)\\
Kaniadakis & $\kappa$ = 0.444 (0.0232) & $\kappa$ = 0.421 (0.0067) & $\kappa$
= 0.398 (0.0059) & $\kappa$ = 0.012 (0.0051)\\
Superstatistical & $\nu=-0.188~(0.0021)$ & $\nu=0.617~(0.0002)$ &
$\nu=1.76~(0.0002)$ & $\nu=3.12~(0.0003)$%
\end{tabular}

\bigskip

\pagebreak

\bigskip

\bigskip\textbf{%
\begin{figure}
[ptb]
\begin{center}
\includegraphics[
natheight=3.512000in,
natwidth=4.433000in,
height=3.5587in,
width=4.4849in
]%
{KAJ2BX00.wmf}%
\end{center}
\end{figure}
}

\bigskip

FIG.1 NNS distributions of resonances in mushroom billiards (histograms),
measure by Friedrich et al. \cite{friedrich} compared with the Tsallis (solid
lines), Kaniadakis (dashed) and superstatistical (dashed-dotted) NNS
distributions. The best-fit parameters are given in Table I.

\bigskip

\pagebreak%

\begin{figure}
[ptb]
\begin{center}
\includegraphics[
natheight=3.512000in,
natwidth=4.433000in,
height=3.5587in,
width=4.4849in
]%
{KAJ2CW01.wmf}%
\end{center}
\end{figure}

FIG.2 NNS distributions of geometrical resonances in random network (dots),
calculated by Gu et al. \cite{gu} compared with the Tsallis (solid lines),
Kaniadakis (dashed) and superstatistical (dashed-dotted) NNS distributions.
The best-fit parameters are given in Table II.

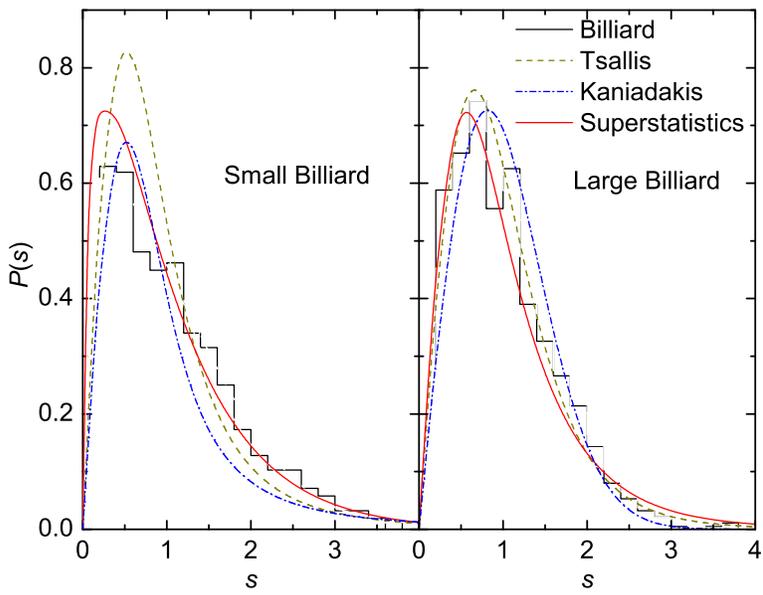
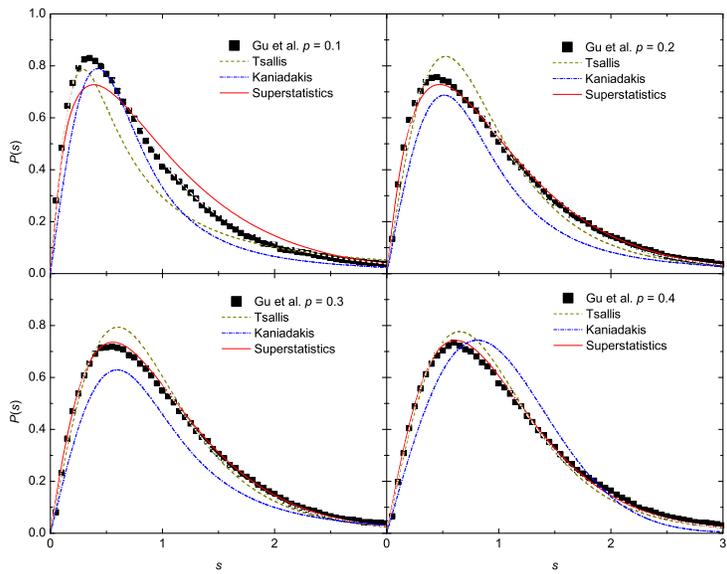
\end{document}